# Enhancing atomic ordering, magnetic and transport properties of Mn$_2$VGa Heusler alloy thin films toward negatively spin-polarized charge injection

Z.H. Li[a], H. Suto[a*], V. Barwal[a], K. Masuda[a], T.T. Sasaki[a], Z.X. Chen[a], H. Tajiri[b], L.S.R. Kumara[b], T. Koganezawa[b], K. Amemiya[c], S. Kokado[d], K. Hono[a], Y. Sakuraba[a]

[a] National Institute for Materials Science, 1-2-1 Sengen, Tsukuba, 305-0047, Japan

[b] Japan Synchrotron Radiation Research Institute, 1-1-1 Kouto, Hyogo, 679-5198, Japan

[c] Institute of Materials Structure Science, High Energy Accelerator Research Organization, 1-1 Oho, Tsukuba, 305-0801, Japan

[d] Graduate School of Integrated Science and Technology, Shizuoka University, 3-5-1 Johoku, Shizuoka, 432-8561, Japan

*Corresponding author: suto.hirofumi@nims.go.jp

**Abstract**

Magnetic materials with negative spin polarization have attracted attention for their potential to increase the design freedom of spintronic devices. This study investigated the effects of off-stoichiometry on the atomic ordering, microstructure, and magneto-transport properties in Mn$_{2+x}$V$_{1-x}$Ga ($x = -0.2, 0, +0.2, +0.4$) Heusler alloy films, which are predicted to have large negative spin polarization derived from a pseudo band gap in the majority spin channel. The Mn$_{2+x}$V$_{1-x}$Ga films epitaxially grown on MgO(001) substrates exhibits variations of *B*2 and *L*2$_1$ order with the Mn concentration. A high-quality *L*2$_1$ ordered film was achieved in the Mn-rich composition ($x = +0.2$) with *B*2 and *L*2$_1$ order parameters of 0.97 and 0.86, respectively, and a saturation magnetization of 1.4 $\mu_B$/f.u, which agrees the Slater-Pauling rule. Scanning transmission electron microscopy observations showed that *B*2 and *L*2$_1$ phases coexist in Mn-poor and stoichiometric films, while the *L*2$_1$ phase is dominant in the Mn-rich film with small amounts of Mn–V and Mn–Ga disorders, as revealed by laboratory and anomalous X-ray diffraction. Combined first-principles calculations and anisotropic magnetoresistance analysis confirm that the addition of excess Mn preserves the high spin polarization by suppressing the formation of detrimental antisites of V atoms occupying Mn sites. Therefore, the Mn-rich composition is promising for negatively spin-polarized charge injection in Mn$_2$VGa-based spintronic applications.

## 1. Introduction

Owing to the unique energy band structure with a bandgap at the Fermi level ($E_F$) of one spin-channel, half-metallic Heusler alloys have attracted considerable attention for potential applications in spintronic devices such as magnetic random-access memories, read sensors and spin-torque oscillators for next-generation ultra-high density hard disk drives [1,2]. Among the Heusler alloy family, numerous studies have been conducted on Co-based $Co_2YZ$ alloys (Y = Mn, Fe, Cr, and Z = Al, Si, Ga, Ge etc.) to enhance their spin-dependent properties at room temperature (RT) because of the theoretically predicted 100% spin polarization and high Curie temperature ($T_C$) [3-9]. On the other hand, Mn-based Heusler alloys ($Mn_2YZ$), which are also expected to exhibit half-metallicity, have not been fully explored as alternative candidates [10-12]. The intrinsically low saturation magnetization of Mn-based alloys can reduce the critical current density for magnetization switching by spin-transfer torque, which may lead to low-energy operation in spintronic devices [13,14].

$Mn_2VGa$ ferrimagnetic Heusler alloy has great potential as a spintronic material due to its sufficiently high $T_C$ of ~ 784 K [10,12,15-19]. Mn and V atoms in the $L2_1$ ordered $Mn_2VGa$ are antiferromagnetically coupled (Mn:1.64 $\mu_B$/f.u., V: −1.2 $\mu_B$/f.u.), resulting in a smaller magnetic moment of ~ 2 $\mu_B$/f.u than those of Co-based alloys; ~ 4−6 $\mu_B$/f.u. More interestingly, $Mn_2VGa$ was predicted to have negative spin polarization ($P < 0$) [20], *i.e.*, the energy band structure has a gap in the majority-spin band at the $E_F$, and the spin direction of the spin-polarized current is opposite to the magnetization direction [21]. Magnetic materials with negative $P$ have been demonstrated to significantly improve the design freedom of spintronic devices [22-24]. For example, giant magnetoresistance devices using FeCr magnetic layer showed inverse magnetoresistance, providing evidence of negative $P$ [25-28]. Inverse magnetoresistance has also been measured in tunneling magnetoresistance devices [29-31]; Klewe *et al*. reported the inverse magnetoresistance for an epitaxial $Mn_2VGa$/MgO/$Co_{70}Fe_{30}$ magnetic tunnel junction, confirming the negative $P$ in the Mn$_2$VGa alloy [31]. However, the magnitude of the observed magnetoresistance ratio (−2%) was much smaller than the theoretical value (−43%). Since the spin polarization of Heusler alloys is closely associated with atomic ordering [32,33], it is important to understand the relationship between negative $P$ and atomic disorder in Mn$_2$VGa in order to improve spin-dependent properties.

The atomic disorder has been reported to impair the spin polarization of the Mn$_2$VGa alloy [20,31]. Theoretical calculations predict that the negative $P$ of Mn$_2$VGa with $B$2 ordered structure (−69%) is lower than that of Mn$_2$VGa with $L$2$_1$ ordered structure (−84%) [31]. The Mn–V disorder can deteriorate the pseudo-gap in the majority spin band of $L$2$_1$ ordered Mn$_2$VGa alloy, leading to the degradation of negative $P$ [31]. Compositional optimization is an efficient way to improve the atomic ordering in Heusler alloys [34-38]. For example, a stoichiometric Co$_2$Fe(Ga$_{0.5}$Ge$_{0.5}$) Heusler alloy film showed a low degree of $L$2$_1$ order; ~ 0.13, which was significantly improved to ~ 0.86 by adding excess Ge to suppress deleterious Co$_{Fe}$ antisite [38], where $X_Y$ denotes that the atomic sites for element $Y$ are replaced with $X$. The calculated band structure of the Ge-rich composition predicts the higher spin polarization than the stoichiometric composition, which is indicated by anisotropic magnetoresistance (AMR) analysis. Based on an extended two-current model developed by Kokado *et al.* [39], the AMR effect has been experimentally demonstrated to be closely related to the spin polarization of various Co-based Heusler alloy films [40-42]. Therefore, an in-depth investigation of the effects of composition tuning along with optimized thermal processing on the atomic ordering, microstructure, and magneto-transport properties of Mn$_2$VGa alloys is required to gain insights into the formation of antisite defects and their relationship with electronic structures.

In this study, we fabricated stoichiometric and off-stoichiometric Mn$_{2+x}$V$_{1-x}$Ga (MVG) alloy thin films with nominal compositions of $x =$ −0.2, 0, +0.2, and +0.4 by the magnetron sputtering. The concentrations of Mn and V were tailored because these two elements mainly determine the ferrimagnetic and spintronic properties of Mn$_2$VGa. Two heat treatments; elevated-temperature deposition and post-deposition annealing were carried out to obtain the optimum thermal processing parameters. The atomic ordering and antisite defects of the prepared MVG films were quantitatively evaluated using a combination of laboratory/synchrotron X-ray diffraction (XRD) analyses. Laboratory XRD (lab-XRD) cannot differentiate the site occupancies of Mn and V due to their close atomic numbers, whereas it can be distinguished by synchrotron XRD. Aberration-corrected scanning transmission electron microscopy (STEM) was performed to examine the uniformity of the films with Mn and V atoms being identified in real space. The revealed ordering and defects were then correlated with the magnetic and transport properties. Based on experimental results and first-principles calculations of electronic structure, the relationship between atomic ordering and spin polarization in MVG films is discussed.

## 2. Experimental and computational methods

Mn$_{2+x}$V$_{1-x}$Ga ($x = -0.2, 0, +0.2, +0.4$) thin films were grown on (001)-oriented MgO single crystalline substrates using an automatic co-sputtering system with Mn$_{55}$Ga$_{45}$ and V targets. The Mn/Ga ratio was controlled by the Ar pressure, and the V content was controlled by the sputtering power. Elevated-temperature deposition and post-deposition annealing were performed on the stoichiometric samples ($x = 0$) at substrate temperatures ($T_{sub}$) of 300–600°C and $T_{ann}$ = 300–600°C for 30 min, respectively. The off- stoichiometric samples ($x = -0.2, +0.2,$ and $+0.4$) were prepared by post annealing at $T_{ann}$ = 500°C and 600°C. All films were capped by 2 nm-thick Al layers after cooling down to RT. **Table 1** summarizes the chemical compositions of Mn$_{2+x}$V$_{1-x}$Ga films in at.% measured using X-ray fluorescence (XRF) analysis calibrated by the standard sample whose composition was analyzed by inductively coupled plasma mass spectrometry. The XRF measurements were repeated five times for each film, and the standard deviations are listed as errors in Table 1. The errors are significantly smaller than the compositional variations between the films, indicating the reliability of the measured compositions. Note that the actual Mn concentration of the Mn-rich films ($x = +0.2$ and $+0.4$) are slightly lower than the nominal concentrations. The deviation may occur because the composition was adjusted by depositing the samples at RT, and the heat treatments caused slight evaporation of Mn atoms.

The crystal structure and atomic ordering of the prepared films were investigated using lab-XRD with a Cu–$K\alpha$ radiation source. The degree of $B2$ and $L2_1$ ordering, $S_{B2}$ and $S_{L2_1}$ were calculated based on $S_{B2} = \sqrt{\frac{I^{obs}_{002}/I^{obs}_{004}}{I^{sim}_{002}/I^{sim}_{004}}}$ and $S_{L2_1} = \sqrt{\frac{I^{obs}_{111}/I^{obs}_{444}}{I^{sim}_{111}/I^{sim}_{444}}}$, where $I^{obs}_{hkl}$ and $I^{sim}_{hkl}$ represent the observed and simulated $hkl$ peak intensities, respectively [43-45]. The observed peak intensity was analyzed by fitting with split pseudo-Voigt function. The XRD pattern simulation was performed using the Visualization for Electronic and Structural Analysis (VESTA) software based on the XRF compositions. Given the close atomic scattering factors of Mn ($f_{Mn}$) and V ($f_V$) in the Cu–$K\alpha$ source, anomalous XRD (AXRD) measurements were performed using synchrotron-radiated X-rays (BL13XU, SPring-8) to evaluate the degree of Mn-V order as the anomalous dispersion term of $f_{Mn}$ significantly changes around the Mn $K$-absorption edge (6.539 keV) [46,47]. X-ray magnetic circular dichroism (XMCD) measurements were performed at the Photon Factory of the High Energy Accelerator Research Organization (BL-7A, KEK) using the total electron yield (TEY) method with a magnetic field of $\pm$ 1 T applied along the incident X-ray beam. The magnetic properties were measured using vibrating sample magnetometer at RT. High-angle annular dark-field scanning transmission electron microscopy (HAADF-STEM) observations, nano-beam electron diffraction (NBED) and energy-dispersive X-ray spectroscopy (EDS) were performed using an FEI Titan G$^2$ 80-200 TEM with a probe aberration corrector operating at 200 kV. Thin foil specimens for STEM observations were prepared by the focused ion beam lift-out technique using an FEI Helios G4 UX.

Table 1 Chemical compositions of the $Mn_{2+x}V_{1-x}Ga$ films measured by XRF analysis.

| Sample | Thermal treatment | Mn (at.%) | V (at.%) | Ga (at.%) | Composition ratio |
|---|---|---|---|---|---|
| $x = 0$ | $T_{sub}$= 600°C | 48.62±0.07 | 26.1±0.05 | 25.28±0.09 | $Mn_{1.94}V_{1.04}Ga_{1.01}$ |
| $x = -0.2$ | $T_{ann}$= 600°C | 44.27±0.09 | 29.61±0.14 | 26.12±0.12 | $Mn_{1.77}V_{1.19}Ga_{1.04}$ |
| $x = 0$ | $T_{ann}$= 600°C | 48.95±0.14 | 25.78±0.08 | 25.27±0.07 | $Mn_{1.96}V_{1.03}Ga_{1.01}$ |
| $x = +0.2$ | $T_{ann}$= 600°C | 52.33±0.07 | 20.56±0.07 | 27.11±0.03 | $Mn_{2.09}V_{0.83}Ga_{1.08}$ |
| $x = +0.4$ | $T_{ann}$= 600°C | 58.25±0.06 | 14.92±0.02 | 26.82±0.05 | $Mn_{2.33}V_{0.6}Ga_{1.07}$ |

The density of states (DOS) of $B2$ and $L2_1$ ordered $Mn_{2+x}V_{1-x}Ga$ were calculated based on the density functional theory and the Korringa-Kohn-Rostoker method implemented in the Akai-KKR software package [48-50]. The generalized gradient approximation was used for the exchange correlation energy [51], and the disordered states were treated within the coherent potential approximation. For all these calculations, we used a lattice constant of $a = 5.905$ Å, which was reported in a previous bulk study [15]. The lattice constant of the films prepared in this study is close to the bulk value as measured by XRD. The Brillouin-zone integration was performed using $16 \times 16 \times 16$ k points, and the imaginary part of the energy was set to 0.001 Ry for the DOS calculations. The spin-orbit interaction was considered in the DOS calculations for analyzing the AMR effect.

For the AMR measurements, MVG films were patterned into strips with dimensions of 2.8 mm × 0.6 mm using photolithography and Ar ion-milling. The electric resistivities, $\rho$, of stripes were measured by a four-terminal method using a physical property measurement system (PPMS DynaCool, Quantum Design) over a temperature range of 10–300 K. A current, $I$, of 500 µA along the longitudinal direction of stripes ($[110]_{MVG}$ or $[100]_{MVG}$), and a magnetic field of 1 T, which is sufficient to saturate the magnetization, rotating within the film plane were applied during the measurements. The angle, $\phi$, dependence of the AMR ratio was defined as $(\rho_{(\phi)} - \rho_\perp)/\rho_\perp \times 100\,\%$, where $\phi$ represent the relative angle between the current and magnetic field directions, and $\rho_{(\phi)}$ ($\rho_\perp$) is the resistivity of films at $\phi$ ($\phi = 90°$).

## 3. Results

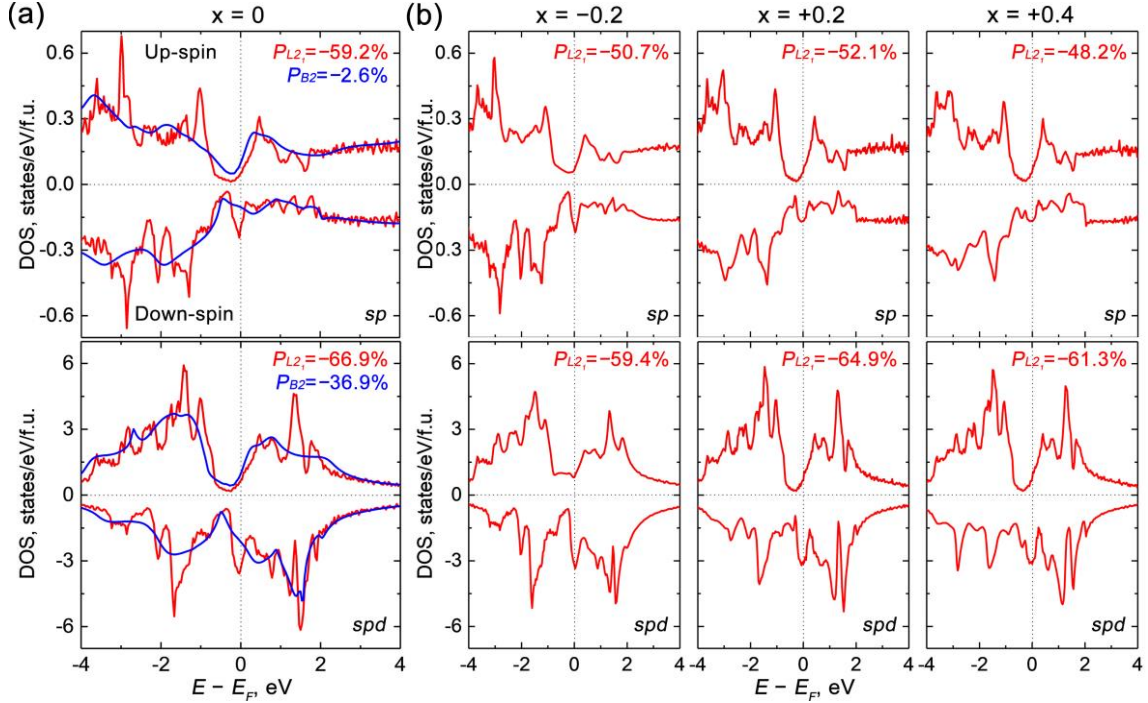

**Fig. 1.** First-principles calculations of the *spd*-total and *sp*-orbital DOSs for $L2_1$ and $B2$ ordered $Mn_{2+x}V_{1-x}Ga$ with (a) $x = 0$ (stoichiometric), and (b) $x = -0.2$, $+0.2$, and $+0.4$ (off-stoichiometric), respectively.

**Figures 1a and b** show the calculated energy dependence of the *sp*-orbital and *spd*-total DOSs for perfect $L2_1$ and $B2$ ordered $Mn_{2+x}V_{1-x}Ga$ ($x = 0, -0.2, +0.2, +0.4$) alloys with nominal compositions. Note that in the calculation model for Mn-poor (Mn-rich) composition, excess V (Mn) atoms occupy the sites of deficient Mn (V) atoms. The DOS of the stoichiometric $Mn_2VGa$ ($x = 0$) exhibits a pseudo-gap around the $E_F$ in the majority-spin band, resulting in negative $P$, **Fig. 1a**. The values of $P$ at the $E_F$ of $L2_1$ ordered state are calculated to be $-59.2\%$ and $-66.9\%$ for the *sp*-orbital and *spd*-total DOSs, respectively. The *sp*-orbital DOS is a representative parameter for the spin polarization of conduction electrons due to the small effective mass of *sp*-electrons. The $B2$ ordered state significantly deteriorates the $P$ value to $-2.6\%$ (*sp*-orbital) and $-36.9\%$ (*spd*-total), indicating that a structure with high $L2_1$ ordering is desirable for achieving high negative spin polarization. In the Mn-poor composition ($x = -0.2$), the gap structure was disturbed by the increased DOS in the majority-spin pseudo-gap, leading to a decrease in $P$ value to −50.7% (*sp*-orbital). On the other hand, the gap structure is maintained in the Mn-rich compositions, $x = +0.2$ and $+0.4$. Despite the maintained pseudo-gap, $P$ becomes smaller than that of the stoichiometric composition, because the $E_F$ exists at the edge of the pseudo-gap and is sensitive to the small change in the DOS. These calculations indicate that excess V atoms in the Mn-site, *i.e.*, $V_{Mn}$ antisites are detrimental to $P$ by inducing the fundamental change in DOS, **Fig. 1b**.

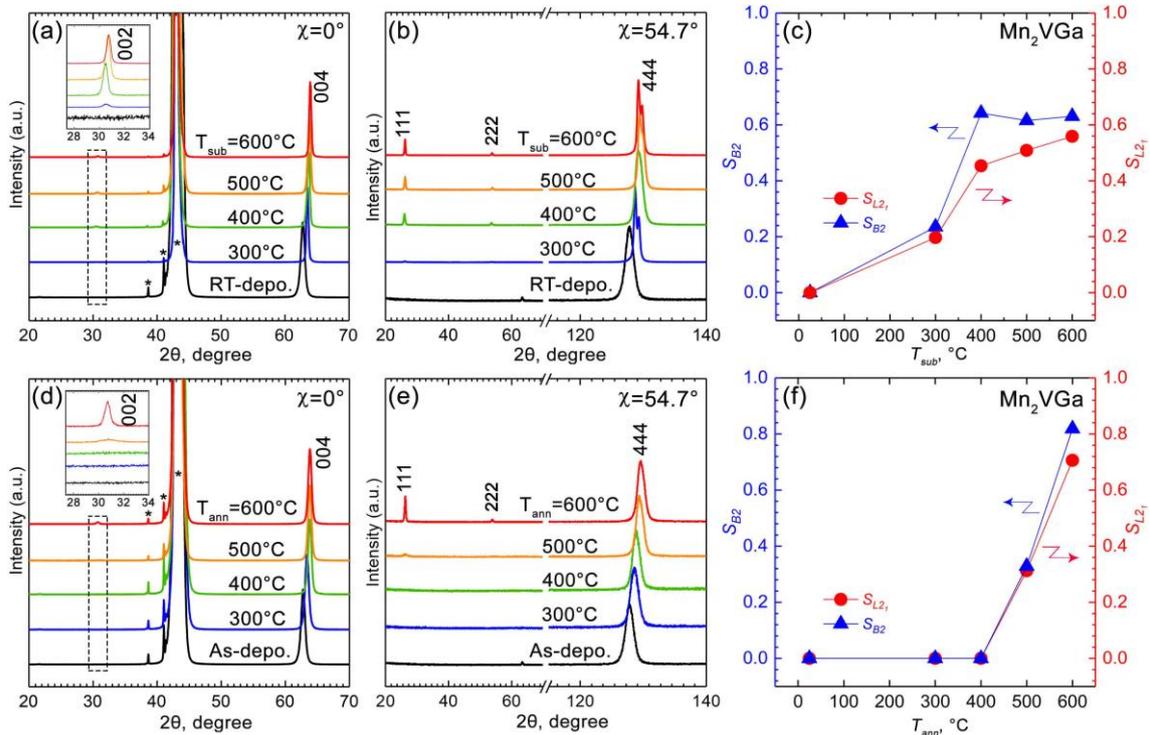

**Fig. 2.** Lab-XRD profiles measured from the out-of-plane ($\chi = 0°$) and $\langle 111 \rangle$ directions ($\chi = 54.7°$), and temperature dependence of $S_{B2}$ and $S_{L2_1}$ for the Mn$_2$VGa films (a–c) deposited at various $T_{sub}$, and (d–f) deposited at RT and post-annealed at various $T_{ann}$. Note that the insets in (a) and (d) shows enlarged (002) peaks, and the peaks labeled with * are from the MgO substrate.

**Figure 2a** shows the lab-XRD profiles of the Mn$_2$VGa ($x = 0$) films deposited at RT and $T_{sub} = 300–600°C$ with scattering vectors set to the out-of-plane ($\chi = 0°$). All films exhibit diffraction peaks only from the {00*l*} plane, indicating 001-oriented epitaxial growth of the Mn$_2$VGa films over the entire range of $T_{sub}$. The 004 peak position shifts toward a higher angle with increasing $T_{sub}$, which suggests lattice shrinkage along the [001] direction. The 002 superlattice peak is not detected at RT and appears as the $T_{sub}$ increased above 300°C, indicating the formation of a *B*2 ordered structure. **Fig. 2b** shows XRD profiles measured along the [111] direction of the Mn$_2$VGa films ($\chi$ = 54.7°). Similar to the 002 peak, the 111 superlattice peak becomes visible at $T_{sub}$ above 300°C, indicating an *L*2$_1$ ordered structure in the samples. The shoulders in the 444 peaks are due to the contributions from Cu *K*$\alpha_1$ and *K*$\alpha_2$ lines with a small difference in wavelength [52]. **Fig. 2c** shows the variations of the degree of *B*2 and *L*2$_1$ order *i.e.*, $S_{B2}$ and $S_{L2_1}$ as a function of $T_{sub}$. The typical errors in the calculated $S_{B2}$ and $S_{L2_1}$ were less than 0.02 as estimated from the errors in the fitting. The value of $S_{B2}$ increases to a maximum of 0.64 at 400°C and remains almost constant from 400 to 600°C. In contrast, $S_{L2_1}$ increases monotonically with increasing $T_{sub}$, reaching a maximum value of 0.56 at 600°C. In the case of post-annealed Mn$_2$VGa films, 002 and 111 peaks are not observed below $T_{ann}$ = 500°C, as shown in **Fig. 2d and e**. However, the $S_{B2}$ and $S_{L2_1}$ values at $T_{ann}$ = 600°C; 0.82 and 0.71, respectively, are even larger than those of the film deposited at $T_{sub}$ = 600°C, indicating that a higher degree of *B*2 and *L*2$_1$ order can be obtained by post-annealing at high temperatures.

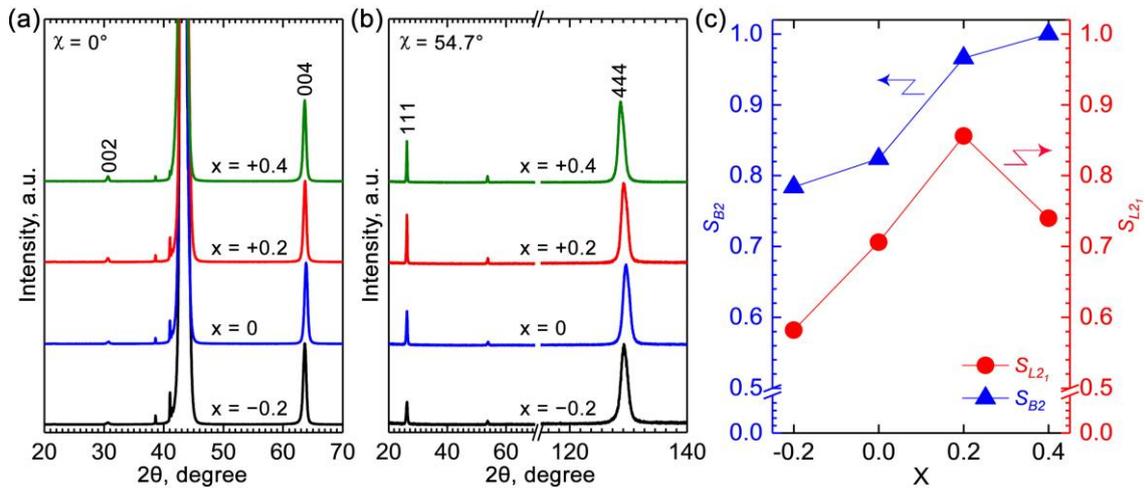

**Fig. 3.** Lab-XRD profiles for MVG films post-annealed at $T_{ann}$ = 600°C measured from the (a) out-of-plane ($\chi$ = 0°) and (b) ⟨111⟩ directions ($\chi$ = 54.7°). (c) The *x* dependence of $S_{B2}$ and $S_{L2_1}$.

**Figure 3a** shows the out-of-plane XRD profiles of Mn$_{2+x}$V$_{1-x}$Ga (*x* = −0.2, 0, +0.2, +0.4) films post-annealed at 600°C. The observed peaks of 002 and 004 indicate the 001-oriented epitaxial growth and the presence of *B*2 ordering in all the compositions. Similarly, the 111 superlattice peak is detected from the ⟨111⟩-direction patterns of all MVG films, **Fig. 3b**, indicating the formation of $L2_1$-ordered structure. **Fig. 3c** shows the variations in $S_{B2}$ and $S_{L2_1}$ as functions of $x$. For the Mn-poor sample ($x = -0.2$), the values of $S_{B2}$ and $S_{L2_1}$ are 0.78 and 0.58, which are lower than those for $x = 0$. In contrast, the values of $S_{B2}$ and $S_{L2_1}$ significantly increase to 0.97 and 0.86, respectively at $x = +0.2$, indicating that the addition of Mn of $x = +0.2$ can achieve near-perfect $S_{B2}$ and high $S_{L2_1}$. Further increasing the Mn concentration to $x = +0.4$ results in a slight increase in $S_{B2}$, but a decrease in $L2_1$ to 0.74.

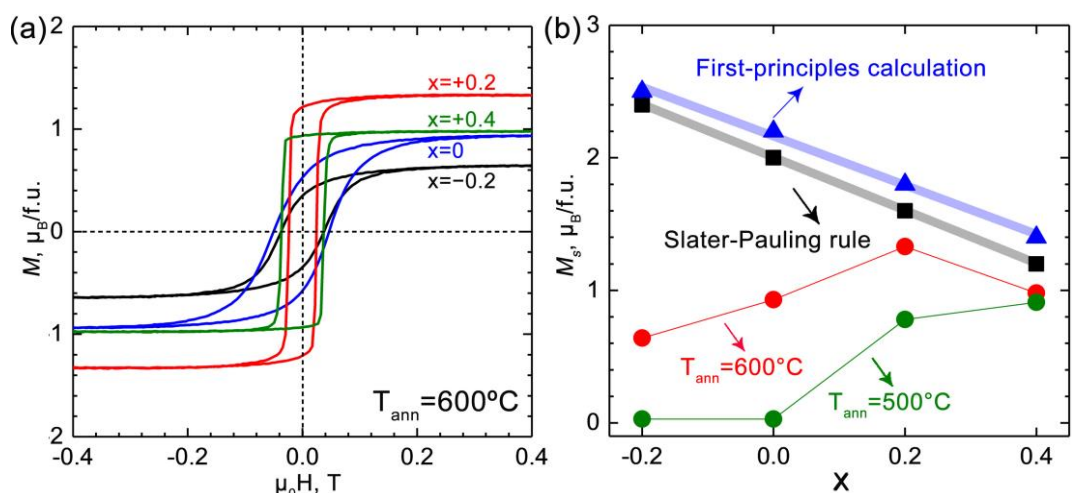

**Fig. 4.** (a) In-plane magnetization curves. (b) Measured and theoretical saturation magnetization, $M_s$ as a function of $x$ for MVG films measured at RT.

**Figure 4a** shows the in-plane magnetization curves for $Mn_{2+x}V_{1-x}Ga$ ($x = -0.2, 0, +0.2,$ and $+0.4$) films at $T_{ann} = 600°C$. All MVG films reach magnetic saturation in the field range of $\pm 0.4$ T. The saturation magnetization, $M_s$ is 0.6 $\mu_B$/f.u for the Mn-poor sample with $x = -0.2$, and increases to 0.9 and 1.3 $\mu_B$/f.u at $x = 0$ and $+0.2$, respectively. Further increasing the Mn concentration to $+0.4$ leads to a decrease in $M_s$. The squareness of the magnetization curves increased for the Mn-rich compositions. Note that the bulk $M_s$ of $Mn_2VGa$ was reported to be 1.7 $\mu_B$/f.u [11]. **Fig. 4b** shows the $x$ dependence of $M_s$ at $T_{ann} = 500$ and $600°C$ together with the values predicted by first-principles calculations and the Slater-Pauling (S-P) rule [53], $M_s = Z_t - 24$, where $Z_t$ is the total number of valence electrons based on the nominal compositions in **Table 1**. The first-principles calculations show a decrease in $M_s$ as $x$ increases, which agrees with the S-P rule. However, the experimental $M_s$ at $T_{ann}$ = 600°C shows the opposite trend from $x$ = −0.2 to +0.2, and the deviation between the experimental and calculation values decreases with increasing $x$, implying the improved atomic ordering for the Mn-rich films. The experimental trend of $x$ = +0.2 and +0.4 follows the first-principles calculations and the S-P rule, reflecting that these two compositions have similarly high ordering. At $T_{ann}$ = 500°C, the Mn-poor and stoichiometric sampled show no magnetization, which is consistent with the XRD results that atomic ordering is poor at this temperature. In contrast, the Mn-rich samples are well magnetized; the $M_s$ value at 500°C for $x$ = +0.2 is smaller than that at 600°C, whereas $x$ = +0.4 has almost the same $M_s$ value at 500 and 600°C. These results indicate that the Mn-rich compositions lower the ordering temperature in addition to the improved atomic ordering. Considering that highest $L2_1$ ordering and $M_s$ close to the theoretical value are achieved in the Mn-rich sample with $x$ = +0.2, we treat the $x$ = +0.2 sample as representative of the Mn-rich compositions. The following detailed characterizations were performed on the samples with $x$ = −0.2, 0, and +0.2.

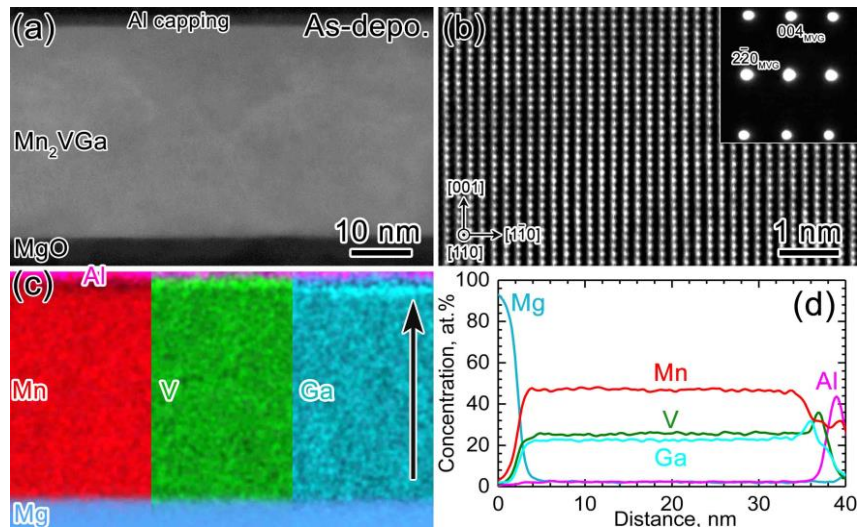

**Fig. 5.** (a) Low-magnification cross-sectional HAADF-STEM image of a MgO sub./Mn$_2$VGa/Al structure. (b) Atomic-resolution HAADF-STEM image and the corresponding NBED pattern of the Mn$_2$VGa layer. (c) EDS elemental maps for Al, Mn, V, Ga, and Mg, respectively, and (d) the corresponding line compositional profile along the direction indicated by the arrow in (c). Note that electron beam is parallel to the [110]$_{Mn2VGa}$.

**Figure 5a** shows a low-magnification cross-sectional HAADF-STEM image of the Mn$_2$VGa/Al films deposited at RT without post annealing. Note that the image was taken along the zone axis of [110]$_{Mn2VGa}$. The imaging contrast within the Mn$_2$VGa layer is nearly homogeneous, indicating uniform growth of the film without pronounced phase separation. The atomic resolution HAADF-STEM image of the as-deposited Mn$_2$VGa shows uniform atomic column intensity, which suggests a random distribution of Mn, V, and Ga atoms, *i.e.*, disordered A2 structure, **Fig. 5b**. Accordingly, only the fundamental A2 reflections of $2\bar{2}0$ and 004 are detected in the NBED pattern. **Fig. 5c** shows the EDS elemental maps of Al, Mn, V, Ga, and Mg. The constituent elements exhibit a uniform distribution within the Mn$_2$VGa layer. The corresponding EDS line compositional profile analyzed along the direction indicated by the arrow in **Fig. 5c** reveals a nearly stoichiometric composition of Mn$_{1.97}$V$_{1.08}$Ga$_{0.95}$, **Fig. 5d**. In addition, slight V and Ga segregation is detected near the Mn$_2$VGa/Al interface.

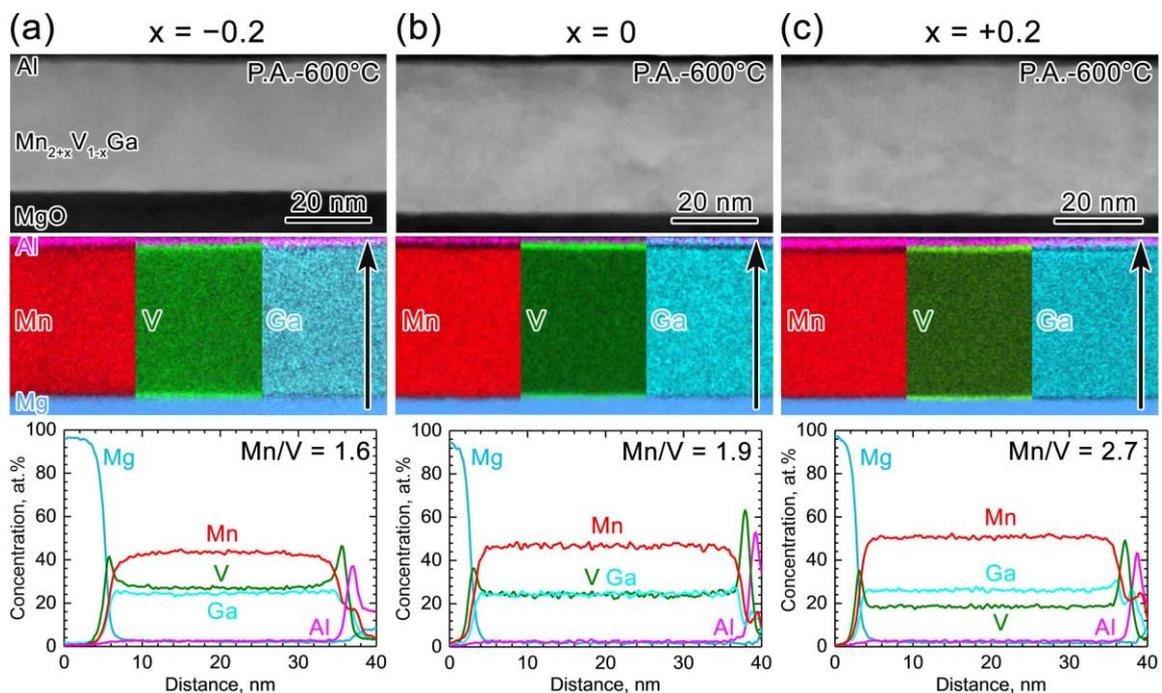

**Fig. 6.** Low-magnification cross-sectional HAADF-STEM images, EDS elemental maps, and compositional line profiles for post-annealed Mn$_{2+x}$V$_{1-x}$Ga films with $x =$ (a) −0.2, (b) 0, and (c) +0.2, respectively.

**Figures 6a–c** show cross-sectional HAADF-STEM images and EDS elemental maps of MgO sub./Mn$_{2+x}$V$_{1-x}$Ga/Al films post-annealed at 600 ºC. All MVG layers show homogeneous

imaging contrast and sharp interfaces with Al capping layers, indicating the uniform growth. Cross-sectional EDS elemental maps and corresponding line compositional profiles show homogeneous distributions of Mn, V, and Ga elements in all MVG layers. The compositions for $x = -0.2$, 0, and +0.2 are measured to be $Mn_{1.83}V_{1.14}Ga_{1.03}$ (Mn/V = 1.6), $Mn_{1.96}V_{1.03}Ga_{1.01}$ (Mn/V = 1.9) and $Mn_{2.13}V_{0.78}Ga_{1.09}$ (Mn/V = 2.7), respectively, which agree well with the XRF results. Compared with the RT-deposited $Mn_2VGa$ film, **Fig. 6**, all post-annealed MVG films show the V segregation at both MgO/MVG and MVG/Al interfaces. Close inspection of the MgO/MVG interface shows that the V segregation is due to the formation of an epitaxial pure V phase with a thickness of ~ 1 nm, **Fig. S1**. Considering the paramagnetic nature of V, the V segregation has little influence on the magnetic properties of the MVG layers.

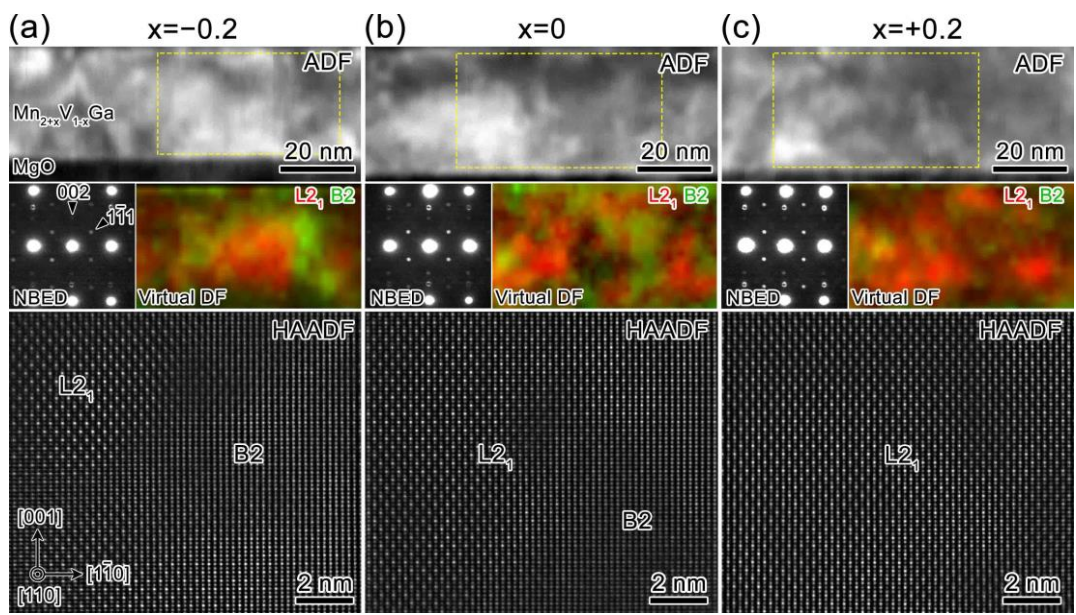

**Fig. 7.** Low-magnification ADF-STEM images, NBED patterns, VDF images of $L2_1$ and $B2$ structures, and atomic resolution HAADF-STEM images for post-annealed $Mn_{2+x}V_{1-x}Ga$ films with $x$ = (a) -0.2, (b) 0, and (c) 0.2, respectively. Note that ADF/HAADF images and NBED patterns were taken along the $[110]_{MVG}$ zone axis, and $L2_1$ and $B2$ phases in the VDF images were reconstructed with a virtual aperture on the 111 and 002 superlattice spots in NBED patterns, respectively.

**Figures 7a–c** show ADF/HAADF-STEM images, NBED patterns, and virtual dark-field (VDF) images of post-annealed $Mn_{2+x}V_{1-x}Ga$ ($x = -0.2$, 0, +0.2) films. The NBED patterns were obtained by scanning the yellow dashed-box areas in the ADF images, and the VDF

images illustrate the intensity distributions of 111 and 002 reflections in the same areas. The $L2_1$ ordered structure was colored in both red and green, while the $B2$ ordered structure was colored in green only. All MVG films show 002 and 111 superlattice spots in the NBED patterns, which confirms the formation of $B2$ and $L2_1$ structures in the whole range of $x$. The intensity of 111 and 002 superlattice reflections increases with increasing $x$, indicating the enhancement of both $B2$ and $L2_1$ ordering by the Mn addition. The VDF images show a co-existence of $B2$ (green) and $L2_1$ (red) structures in the Mn-poor ($x = -0.2$) and stoichiometric ($x = 0$) samples, **Figs. 7a and b**, which are further revealed by atomic resolution HAADF-STEM images in different contrast modulations along the [001] direction; the $L2_1$ structure shows a periodic arrangement of brightly imaged Ga-rich columns separated by three darkly imaged Mn/V columns, while only one-column interval is observed between two neighboring Ga-rich columns in the $B2$ structure. In contrast, the Mn-rich sample ($x = +0.2$) shows the predominance of the $L2_1$ phase, **Fig. 7c**, suggesting the improved atomic ordering due to the excess Mn addition.

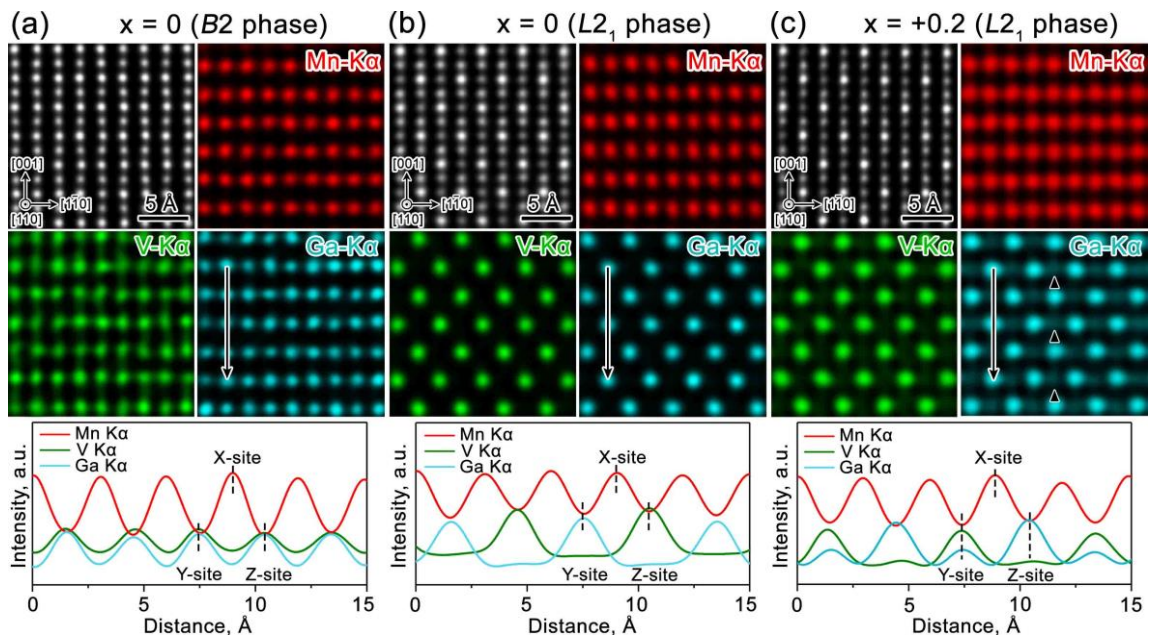

**Fig. 8.** Atomic resolution HAADF-STEM images, EDS elemental maps, and compositional line profiles of Mn, V, and Ga for post-annealed $Mn_{2+x}V_{1-x}Ga$ films with (a,b) $x = 0$ and (c) $x = +0.2$ acquired along $[110]_{MVG}$ zone axis.

**Figures 8a-c** show magnified HAADF-STEM images and the corresponding atomic-resolution EDS elemental maps obtained from stoichiometric ($x = 0$) and Mn-rich ($x = +0.2$) samples. Note that EDS elemental maps were processed using radial Wiener filtering to enhance the imaging contrast and signal-to-noise ratio of elemental line profiles. The site occupancy of each element in ideal and disordered $B2$/$L2_1$ structural models is summarized in **Table 2**. For the stoichiometric sample ($x = 0$), the $B2$ phase shows alternating Mn and V-Ga atomic layers along the [001] direction, **Fig. 8a**. The same elemental occupancy in the V-Ga layer suggests an intermixing of V and Ga atoms (Model I), as evidenced by the EDS line composition profile along the arrow direction. On the other hand, Mn, V, and Ga elements in the $L2_1$ phase periodically occupy individual atomic columns, **Fig. 8b**, which is consistent with the ideal $L2_1$ structure in Model II. In contrast, the Mn-rich sample ($x = +0.2$) shows faint Ga signals (indicated by arrows) in the V sites, respectively, in addition to an ordered atomic arrangement similar to the $L2_1$ phase in the stoichiometric sample. Combined with EDS line scan analysis, weak Ga intensity peaks are clearly observed in the original V sites (Y-site), indicating the intermixing between V and Ga atoms, which is consistent with the ideal structure in model III. In addition, Mn is also expected to partially occupy the V sites based on the XRF composition of $Mn_{2.09}V_{0.8}Ga_{1.08}$, however, no obvious Mn signal or intensity peak is detected from EDS results.

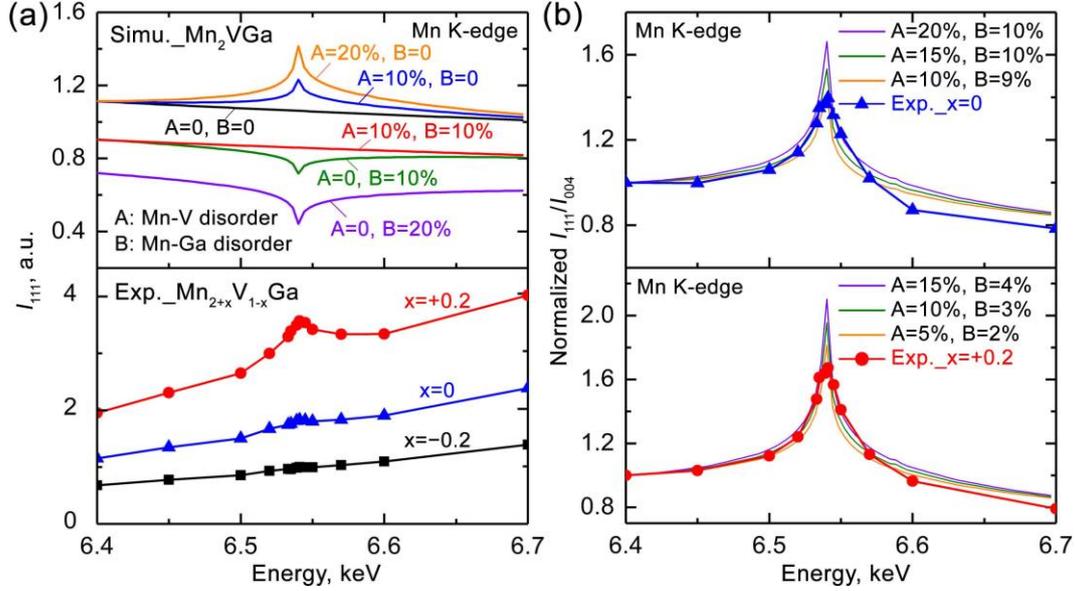

**Fig. 9.** Simulated and experimental X-ray energy dependence of (a) $I_{111}$ for post-annealed MVG films and $L2_1$-structure $Mn_2VGa$, and (b) normalized $I_{111}/I_{004}$ for $x = 0$ and $+0.2$ samples with various Mn-V (A) and Mn-Ga (B) disorders, respectively, near the Mn $K$-edge. Note that the increased background slope with increasing $x$ in (a) is due to the X-ray absorption by air.

**Table 2** Site occupancy of Mn, V, and Ga for ideal $L2_1$, $B2$, and disordered structures in stoichiometric ($x = 0$, $Mn_{1.96}V_{1.03}Ga_{1.01}$) and Mn-rich ($x = +0.2$, $Mn_{2.09}V_{0.83}Ga_{1.08}$) samples.

| Model No. | Structure | X-site (0, 0, 0), (1/2, 1/2, 1/2) | | | Y-site (1/4, 1/4, 1/4) | | | Z-site (3/4, 3/4, 3/4) | | |
|---|---|---|---|---|---|---|---|---|---|---|
| | | Mn | V | Ga | Mn | V | Ga | Mn | V | Ga |
| I | $x = 0$ (ideal $B2$) | 96 | 3 | 1 | 0 | 50 | 50 | 0 | 50 | 50 |
| II | $x = 0$ (ideal $L2_1$) | 96 | 3 | 1 | 0 | 100 | 0 | 0 | 0 | 100 |
| III | $x = +0.2$ (ideal $L2_1$) | 100 | 0 | 0 | 9 | 83 | 8 | 0 | 0 | 100 |
| IV | $x = 0$ (with disorders) | 81 | 10 | 9 | 10 | 80 | 10 | 9 | 10 | 81 |
| V | $x = +0.2$ (with disorders) | 96.5 | 2.5 | 1 | 14 | 73 | 13 | 2 | 5 | 93 |

Considering that the site occupancy revealed by atomic STEM-EDS maps is localized, the average information of atomic ordering in MVG films was further determined by combined lab-XRD and AXRD analysis. **Figure 9a** shows the simulated and experimental X-ray energy dependence of $I_{111}$ for the MVG around the Mn $K$-absorption edge. The $I_{111}$ profiles of $Mn_2VGa$ ($x = 0$) were simulated with various fractions of Mn-V (A) and Mn-Ga (B) disorders based on $F_{111}^{L2_1} = |\{Af_{Mn}+(1-A)f_V\} - \{Bf_{Mn}+(1-B)f_{Ga}\}|$, where $f_x$ is the atomic scattering factor of element *x*. The simulated $I_{111}$ profile exhibits a flat shape for the same amount of Mn-V and Mn-Ga disorders (A = B = 0 and 10%) due to the absence of $f_{Mn}$ in $F_{111}^{L2_1}$, whereas an increase in Mn-V or Mn-Ga disorder changes the profile to be convex or concave, respectively. A comparison between experimental and simulated results suggests that stoichiometric (*x* = 0) and Mn-poor (*x* = −0.2) samples have almost the same amount of Mn-V and Mn-Ga disorders, respectively due to the flat-shape $I_{111}$ profiles. However, the disorder in the Mn-rich sample (*x* = +0.2) cannot be determined solely from the shape of the $I_{111}$ profile because even the ideal $L2_1$ ordered $Mn_{2.09}V_{0.83}Ga_{1.08}$ (A = B = 0) shows a peak near the Mn *K*-edge as the excess Mn atoms (9%) intrinsically occupying the V sites. Instead, we simulated the X-ray energy dependence of normalized $I_{111}/I_{004}$ for stoichiometric and Mn-rich samples, as shown in **Fig. 9b**. In the simulation of the stoichiometric sample, A and B were set to 20% and 10%, 15% and 10%, and 10% and 9%, respectively, to reproduce the $S_{B2}$ obtained by lab-XRD. The stoichiometric sample (*x* = 0) exhibits well-matched shapes and intensity between the simulated and experimental profiles by inducing nearly the same amount of Mn-V (~ 10%) and Mn-Ga (~ 9%) disorders, which agrees with the inference from the $I_{111}$ simulations. Recalling $S_{L2_1}$ = 0.71 and $f_{Mn} \approx f_V$ from lab-XRD, **Fig. 3c**, the V-Ga disorder (C) is calculated to be ~ 10% based on $|(f_V - f_{Ga})|S_{L2_1} = |\{(0.1f_{Mn}+(0.9-C)f_V+Cf_{Ga})\} - \{(0.09f_{Mn}+Cf_V+(0.91-C)f_{Ga})\}|$. In contrast, the Mn-rich sample (*x* = +0.2) has much lower disorders of ~ 5% Mn-V and 2% Mn-Ga from the AXRD analysis, as well as ~ 5% V-Ga disorder calculated by XRD. The average site-occupancy models for stoichiometric and Mn-rich samples are summarized in **Table 2**, denoted as Models IV and V, respectively.

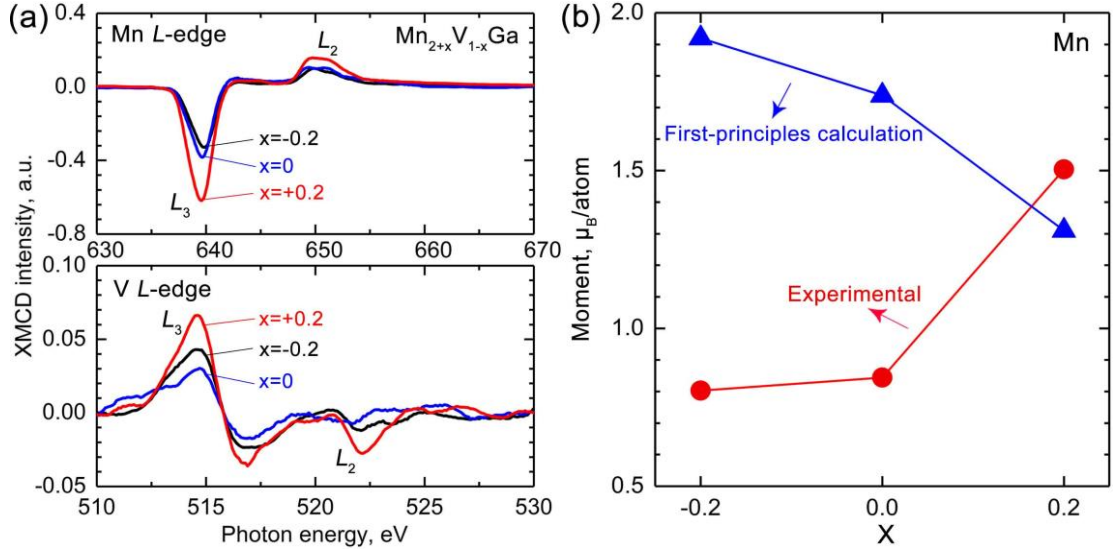

**Fig. 10.** (a) XMCD spectra around Mn and V $L_{3,2}$-edges, and (b) experimental magnetic moments of Mn as a function of $x$ for MVG films measured at RT. Note that the calculated Mn moments based on the $L2_1$-order structural model are also shown for reference.

**Figure 10a** shows the XMCD spectra at the $L$-absorption edges of Mn and V obtained from $Mn_{2+x}V_{1-x}Ga$ ($x = -0.2, 0, +0.2$) films. Note that the probing depth of the TEY method is limited to ~ 2-3 nm from the surface [54], the XMCD results, especially the V moments, are significantly affected by the V segregation on the surface of annealed MVG films, **Fig. 6**. For all films, Mn shows a negative signal at the $L_3$ edge and a positive signal at the $L_2$ edge that are opposite to those of V, confirming the antiferromagnetic coupling between Mn and V sites [10,31]. The signal intensity at the Mn $L_{3,2}$-edges increases with increasing the $x$, which suggests an increase in the Mn moment. The V $L$-edge intensity does not vary monotonically with $x$ due to the surface V segregation. **Fig. 10b** shows the $x$ dependence of the Mn moment determined by the sum rule analysis [55], and the theoretical calculations based on $L2_1$-orderd structure are also provided for reference. The experimental Mn moment slightly increases from 0.8 $\mu_B$/atom at $x = -0.2$ to 0.84 $\mu_B$/atom at $x = 0$, and then significantly rises to 1.5 $\mu_B$/atom with $x$ increasing to +0.2. This trend is consistent with that of $M_s$, supporting the highest of atomic order for the Mn-rich sample. In contrast, a monotonic decrease of the calculated Mn moment with increasing the *x* is attributed to the antiferromagnetic coupling of increased Mn atoms at the V-site with those at the Mn-site, which partially counteracts the Mn moment.

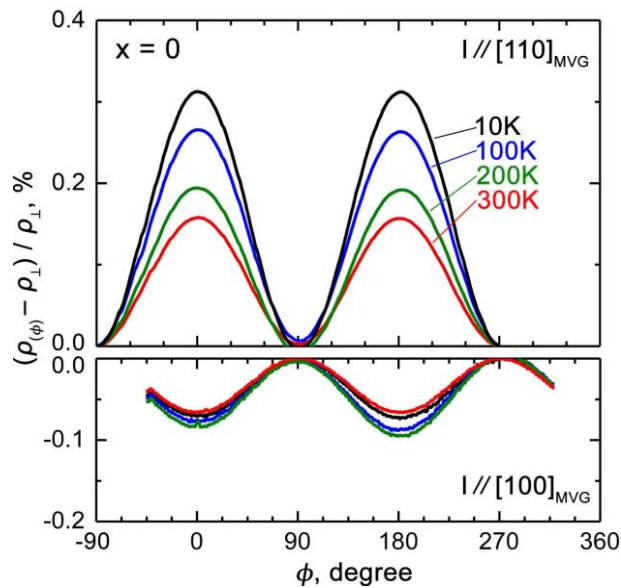

**Fig. 11.** The AMR ratio of the stoichiometric sample ($x = 0$) as a function of $\phi$. Note that the current directions were aligned along $[110]_{MVG}$ and $[100]_{MVG}$, respectively, and the measurement temperature varied from 10 to 300 K.

We performed the AMR analysis to confirm the predictions obtained by the DOS calculations from the viewpoint of transport properties. **Figures 11** shows the $\phi$ dependence of the AMR ratio for the stoichiometric sample ($x = 0$) measured from 10 to 300 K. The composition dependence of the AMR ratio for $Mn_{2+x}V_{1-x}Ga$ ($x = -0.2, 0, +0.2$) films are provided in the Supplementary Materials, **Fig. S.2**. The sign of AMR ratio for $Mn_2VGa$ is positive in the current direction along the $[110]_{MVG}$ and negative along $[100]_{MVG}$, as opposed to those for $Co_2MnGa$, which shows positive and negative values for $I \parallel [110]$ and $I \parallel [100]$, respectively [56]. The AMR values of $Mn_2VGa$ lie between $-0.1\%$ and $+0.3\%$, which is similar to the reported $Co_2FeZ$ and $Co_2MnZ$ with Z = (Al, Si, Ge, Ga) Heusler alloy films with AMR values typically well below 1% [38,40-42].

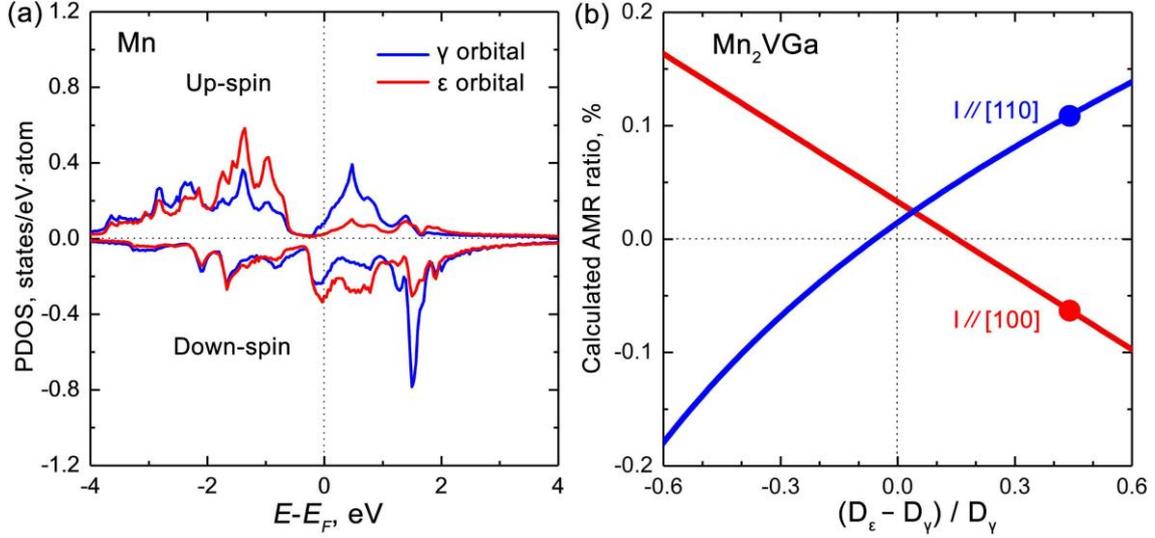

**Fig. 12.** (a) Partial DOS of γ and ε orbitals of Mn $d$ state, and (b) calculated AMR ratio for $L2_1$-ordered Mn$_2$VGa with the current directions aligned along [110]$_{MVG}$ and [100]$_{MVG}$, respectively. Solid circles in (b) indicate the calculated AMR ratios at the $E_F$.

The AMR ratios for the $L2_1$ ordered Mn$_2$VGa were theoretically analyzed by combining first-principles calculations and the developed Kokado's theory [40], which considers the $s$–$d$ scattering of conduction electrons from $s$ to $d$ orbitals under a two-current model, **Figure S.3 and S.4**. **Figure 12a** shows the partial $d$-DOS of γ and ε orbitals of Mn atoms in $L2_1$ ordered Mn$_2$VGa. Here, only the Mn contribution is used due to the dominant spin-down $d$-DOS at $E_F$. According to the expressions given in Eqs. (S1) to (S4), the AMR ratios for two current directions, *i.e.*, [110]$_{MVG}$ and [100]$_{MVG}$ can be described as a function of $\left(D_{\varepsilon,-}^{(d)} - D_{\gamma,-}^{(d)}\right)/D_{\gamma,-}^{(d)}$, where $D_{\varepsilon,-}^{(d)}$ and $D_{\gamma,-}^{(d)}$ are the spin-down $d$-DOS of ε- and γ-orbitals, respectively, **Fig. 12b**. Based on the value of $\left(D_{\varepsilon,-}^{(d)} - D_{\gamma,-}^{(d)}\right)/D_{\gamma,-}^{(d)}$ at the $E_F$ for Mn$_2$VGa; ~ 0.435, the AMR ratio is calculated to be positive (negative) for the [110]$_{MVG}$ ([100]$_{MVG}$), which agrees well with the experimental results (Fig. 11), indicating that the sign reversal of the AMR ratio by the current direction arises from the crystal field of MVG films affecting the $s$–$d$ scattering process. The consistency between the experimental and calculated AMR results also demonstrates the rationality of the DOS calculations for predicting the $P$ of MVG films.

## 4. Discussion

The present work reports an approach to achieve the high $B2$ and $L2_1$ ordering and saturation magnetization, $M_s$ close to the theoretical value in the Mn$_2$VGa-based Heusler alloy films. The degree of $B2$ and $L2_1$ order and $M_s$ of the MVG films were significantly improved by the addition of excess Mn. The Mn-rich sample ($x$ = +0.2) at 600°C post-deposition annealing exhibits a predominant $L2_1$ phase with the highest values of $S_{B2}$ and $S_{L2_1}$, 0.97 and 0.86, respectively, and a maximum $M_s$ of 1.4 $\mu_B$/f.u, which is close to the theoretical value. In addition, the excess Mn addition has no detrimental effect on spin polarization as predicted by the DOS calculations. Therefore, the Mn-rich composition with a high degree of $L2_1$ order is promising for MVG-based spintronic devices with excellent spin-dependent properties.

The stoichiometric Mn$_2$VGa sample shows improved $B2$ and $L2_1$ ordering with increasing annealing temperature, **Figures 2c and f**. The values of $S_{B2}$ and $S_{L2_1}$ by substrate-annealing are higher than those by post-annealing from 300 to 500°C, which is due to the enhanced mobility of adsorbed atoms at elevated substrate temperatures [57,58]. The excess Mn addition further improves the $S_{L2_1}$ of the MVG films, **Fig. 3c**. The large $S_{L2_1}$; ~ 0.86 for the Mn-rich sample at 600°C post-deposition annealing is mainly due to the predominant formation of the $L2_1$ phase, **Fig. 7c**, whereas the stoichiometric sample shows the co-existence of $B2$ and $L2_1$ phases, **Fig. 8c**. Because the order−disorder transition temperature from the $L2_1$ to the $B2$ phase, $T_t^{L2_1/B2}$, in Mn-based Heusler alloys tends to increase with increasing the number of valence electrons [59,60], the Mn-rich sample with $Z_t$= 22.02 is expected to have a higher $T_t^{L2_1/B2}$, *i.e.*, a higher driving force for the $L2_1$-order than the stoichiometric sample, $Z_t$= 21.9. In addition, atomic disorders are reduced by the addition of excess Mn, **Fig. 9**. The decrease in Mn-V and Mn-Ga disorders with increasing Mn concentration is attributed to the preferentially occupation of Mn atoms in the Mn-site, resulting in lower V and Ga occupancy, **Table 2**. The addition of excess Mn can also reduce V-Ga disorders by increasing the chemical potential between V- and Ga-sites [61]. Therefore, the improved $B2$ and $L2_1$ ordering by the excess Mn addition is attributed to the increase in $T_t^{L2_1/B2}$ and the suppression of chemical disorders.

The $M_s$ of $Mn_{2+x}V_{1-x}Ga$ films at $T_{ann}$ = 600ºC increases with increasing Mn concentration, from $x$ = −0.2 to +0.2, **Figure 4a**. However, the trend of $M_s$ varying with Mn concentration contradicts the Slater-Pauling rule and first-principles calculations, **Figure 4b**, suggesting that factors other than the valence electron count may be responsible for the $M_s$ of $Mn_{2+x}V_{1-x}Ga$ films. Considering the antiferromagnetic coupling between Mn- and V-site atoms, **Fig. 10**, the increase in $M_s$ is a result of competition arising from the simultaneously increased magnetic moments of Mn and V with increasing $x$. Because the atomic disorder strongly alters the magnetic interaction between the nearest neighboring atoms [62], the increase in the Mn moment can be explained by decreased Mn–V and Mn–Ga disorders, leading to reduced antiferromagnetic coupling between the Mn atoms of original Mn-site and both $Mn_V$ and $Mn_{Ga}$ antisites. Likewise, the large deviations between the experimental and calculated values for $x$ = −0.2 and 0 can be attributed to a large amount of Mn–V and Mn–Ga disorders formation in the Mn-poor and stoichiometric samples, **Fig. 9**. For Mn-rich samples ($x$ = +0.2 and +0.4), the experimental $M_s$ trend agreed the trend of the Slater-Pauling rule and first-principles calculations, reflecting that these two compositions have similarly high ordering. At $T_{ann}$ = 500ºC, the stoichiometric and Mn-poor samples ($x$ = 0, and −0.2) show no magnetization, In contrast, the Mn-rich samples ($x$ = +0.2, and +0.4) show magnetization, indicating that the Mn-rich compositions can not only improve the ordering but also lower the ordering temperature. In addition to the structural disorder, other factors such as secondary phase and finite temperature effects may also influence the accuracy of the theoretical calculations. Nevertheless, since the variation of $M_s$ is small below the Curie temperature, ~ 784 K and no secondary phase was observed by TEM analysis, the deviation between experimental and calculated $M_s$ is mainly attributed to the atomic disorder in the MVG films.

The DOS calculations for the Mn$_2$VGa revealed that the spin polarization of the $L2_1$ ordered structure is much higher than that of the $B2$ ordered structure, **Figure 1**, indicating that atomic disorder can significantly deteriorate the spin polarization by smearing the band dispersion [45,63]. Compared with the stoichiometric composition ($x = 0$), the pseudo-gap around the $E_F$ is preserved in the Mn-rich composition ($x = +0.2$) but is degraded in the Mn-poor composition ($x = -0.2$), suggesting that the formation of V$_{Mn}$ rather than Mn$_V$ antisites is detrimental to spin polarization. This finding is in contrast to that reported for Co-based Heusler alloy (Co$_2$YZ), where Co$_Y$ rather than Y$_{Co}$ destroys the half-metallic gap, leading to a substantial decrease in spin polarization [4,38,64]. HAADF-STEM observations showed that the $L2_1$ phase increases with increasing Mn concentration, and dominates the Mn-rich sample, **Fig. 7**. Based on the lab-XRD and AXRD analyses, **Figs. 3 and 9**, the average amount of V$_{Mn}$ antisites in the Mn-poor and stoichiometric samples is determined to be ~ 10%, which is four times larger than that in the Mn-rich sample; ~ 2.5%. These results imply that the Mn-rich composition is superior for achieving higher spin polarization because excess Mn addition can readily lead to a high degree of $L2_1$ order while inhibiting the formation of V$_{Mn}$ antisites. Furthermore, the sign reversal of the AMR ratio by the current direction in the stoichiometric sample is consistent with the theoretical calculations, **Figs. 11 and 12**, indicating that the DOS predictions obtained from first-principles calculations is reasonable for the fabricated MVG films.

## 5. Summary

This study has clarified the effects of off-stoichiometry on the atomic ordering, magnetic and transport properties for Mn$_{2+x}$V$_{1-x}$Ga ($x = -0.2, 0, +0.2, +0.4$) films using a combination of (A)XRD, HAADF-STEM, and first-principles calculations. A high degree of $B2$ and $L2_1$ order in the Mn-rich sample ($x = +0.2$) leads to $M_s$ close to the theoretical value, which indicates that the Mn-rich composition is promising for potential spintronic applications. The main conclusions are as follows:

1. The addition of Mn significantly improves $S_{B2}$ and $S_{L2_1}$ of MVG films from 0.78 and 0.58 ($x = -0.2$) to 0.97 and 0.86 ($x = +0.2$), respectively, which is attributed to the predominant formation of the $L2_1$ phase and the suppression of atomic disorder.

2. The increase in $M_s$ with increasing Mn concentration is associated with the competition arising from the simultaneous increase in magnetic moments of Mn and V. The decreased Mn–V and Mn–Ga disorders reduces the antiferromagnetic coupling between the Mn- and V-site atoms, leading to the increase in the Mn and V moments.

3. First-principles calculations indicate that the Mn-rich composition with the $L2_1$ structure exhibits high spin polarization because of the inhibition of deleterious $V_{Mn}$ antisites by the excess Mn addition.

**Declaration of Competing Interest**

The authors declare that they have no known competing financial interests or personal relationships that could have influenced the work reported in this paper.

**Acknowledgements**

This work was partially supported by Advanced Storage Research Consortium (ASRC), MEXT Program: Data Creation and Utilization-Type Material Research and Development Project Grant Number JPMXP1122715503, and JSPS KAKENHI Grant No. 21K20434, No. 23K03934, and No. 19K05249. The experiments were performed under the approval of the Photon Factory Program Advisory Committee (No. 2019S2-003) and the approval of the Japan Synchrotron Radiation Research Institute (JASRI) (No. 2023A1563) The authors thank M. Inoue and Dr. G.Z. Xing of NIMS for technical support and discussion.

**Supplementary Materials**

Supplementary materials associated with this article can be found in the online version.

**References**

[1] A. Hirohata, D.C. Lloyd, Heusler alloys for metal spintronics, MRS Bull. 47 (2022) 593-599.

[2] S. Tavares, K. Yang, M.A. Meyers, Heusler alloys: Past, properties, new alloys, and prospects, Prog. Mater. Sci. 132 (2023) 101017.

[3] T. Ishikawa, H. Liu, T. Taira, K. Matsuda, T. Uemura, M. Yamamoto, Influence of film composition in $Co_2MnSi$ electrodes on tunnel magnetoresistance characteristics of $Co_2MnSi/MgO/Co_2MnSi$ magnetic tunnel junctions, Appl. Phys. Lett. 95 (2009) 232512.

[4] H. Liu, Y. Honda, T. Taira, K. Matsuda, M. Arita, T. Uemura, M. Yamamoto, Giant tunneling magnetoresistance in epitaxial $Co_2MnSi/MgO/Co_2MnSi$ magnetic tunnel junctions by half-metallicity of $Co_2MnSi$ and coherent tunneling, Appl. Phys. Lett. 101 (2012) 132418.

[5] Y. Sakuraba, K. Izumi, T. Iwase, S. Bosu, K. Saito, K. Takanashi, Y. Miura, K. Futatsukawa, K. Abe, M. Shirai, Mechanism of large magnetoresistance in $Co_2MnSi/Ag/Co_2MnSi$ devices with current perpendicular to the plane, Phys. Rev. B 82 (2010) 094444.

[6] Y. Sakuraba, M. Ueda, Y. Miura, K. Sato, S. Bosu, K. Saito, M. Shirai, T.J. Konno, K. Takanashi, Extensive study of giant magnetoresistance properties in half-metallic $Co_2(Fe,Mn)Si$-based devices, Appl. Phys. Lett. 101 (2012) 252408.

[7] H. Narisawa, T. Kubota, K. Takanashi, Current perpendicular to film plane type giant magnetoresistance effect using a Ag-Mg spacer and $Co_2Fe_{0.4}Mn_{0.6}Si$ Heusler alloy electrodes, Appl. Phys. Express 8 (2015) 063008.

[8] S. Li, Y.K. Takahashi, T. Furubayashi, K. Hono, Enhancement of giant magnetoresistance by $L2_1$ ordering in $Co_2Fe(Ge_{0.5}Ga_{0.5})$ Heusler alloy current-perpendicular-to-plane pseudo spin valves, Appl. Phys. Lett. 103 (2013) 042405.
26